\documentclass[10pt]{article}
\usepackage{fancyhdr}
\usepackage{extramarks}
\usepackage{amsmath}
\usepackage{amsthm}
\usepackage{amsfonts}
\usepackage{siunitx}
\usepackage{tikz}
\usepackage[plain]{algorithm}
\usepackage{algpseudocode}
\usepackage{multirow}
\usepackage{booktabs}
\usepackage{palatino}
\usepackage{graphicx}
\usepackage{subfigure}
\usepackage[colorlinks,linkcolor=blue,anchorcolor=blue,citecolor=blue,urlcolor=blue]{hyperref}
\usepackage{amsmath,bm}
\usepackage{booktabs}
\usepackage{mathtools}
\usepackage{amssymb}
\usepackage{caption}
\usepackage{capt-of}
\usepackage{mciteplus}
\usepackage{cite}
\usepackage{mathrsfs}
\usepackage[title,titletoc,toc]{appendix}
\usepackage{xr}
\usepackage{parskip}
\usepackage{soul}
\usepackage{textcomp}
\usepackage[colaction]{multicol}
\usepackage[switch]{lineno}
\usepackage{lipsum}
\usepackage{etoolbox}
\usepackage{longtable}
\usepackage{array}
\usepackage{tablefootnote}
\usepackage{ragged2e}
\usepackage{makecell}
\usepackage{authblk}

\newcolumntype{C}[1]{>{\centering\arraybackslash}p{#1}}
\usetikzlibrary{automata,positioning}
\usetikzlibrary{automata,positioning}

\definecolor{spartangreen}{RGB}{24,69,59}
\definecolor{spartanlightgreen}{RGB}{105,168,105}
\definecolor{spartangray}{RGB}{148,153,140}
\definecolor{spartanlightgray}{RGB}{201,204,198}
\definecolor{spartantan}{RGB}{162,144,87}
\definecolor{spartancream}{RGB}{228,232,222}
\definecolor{spartanheadertop}{RGB}{13,38,31}
\definecolor{nyuultraviolet}{RGB}{87,6,140}
\definecolor{nyuultraviolet}{RGB}{137,0,225}

\newcommand\blfootnote[1]{%
  \begingroup
  \renewcommand\thefootnote{}\footnote{#1}%
  \addtocounter{footnote}{-1}%
  \endgroup
}

\usepackage{tikz,xcolor,hyperref}
\definecolor{lime}{HTML}{A6CE39}
\DeclareRobustCommand{\orcidicon}{%
	\begin{tikzpicture}
	\draw[lime, fill=lime] (0,0) 
	circle [radius=0.16] 
	node[white] {{\fontfamily{qag}\selectfont \tiny ID}};
	\draw[white, fill=white] (-0.0625,0.095) 
	circle [radius=0.007];
	\end{tikzpicture}
	\hspace{-2mm}
}
\foreach \x in {A, ..., Z}{%
	\expandafter\xdef\csname orcid\x\endcsname{\noexpand\href{https://orcid.org/\csname orcidauthor\x\endcsname}{\noexpand\orcidicon}}
}

\begin{document}
\title{VARIANT: Web Server for Decoding and Analyzing Viral Mutations at Genome and Protein Levels}
\author[1]{Rui Wang\orcidA}
\author[2]{Xuhang Dai\orcidE}
\author[1]{Xin Cao\orcidG}
\author[3,$\dagger$]{Changchuan Yin\orcidF}
\author[1,2,4,5,$\dagger$]{Tamar Schlick\orcidB}
\author[6,7,8,$\dagger$]{Guo-Wei Wei\orcidC}

\affil[1]{Simons Center for Computational Physical Chemistry, New York University, New York, NY 10003, USA} 
\affil[2]{Department of Chemistry, New York University, New York, NY 10003, USA}
\affil[3]{Department of Mathematics, Statistics, and Computer Science, University of Illinois at Chicago, Chicago, IL, USA}
\affil[4]{Courant Institute of Mathematical Sciences, New York University, New York, NY 10012, USA}
\affil[5]{New York University-East China Normal University Center for Computational Chemistry, New York University Shanghai, Shanghai 200122, China}
\affil[6]{Department of Mathematics, Michigan State University, East Lansing, MI 48824, USA}
\affil[7]{Department of Mathematics, 		University of Georgia, Athens, GA 30602, USA.}
\affil[8]{Department of Biochemistry and Molecular Biology, University of Georgia, Athens, GA 30602, USA}
 
% \date{\today} % Date for the report

\maketitle

\begin{abstract}
    A comprehensive analysis of viral mutations is essential for understanding viral evolution, disease epidemiology, diagnosis, drug resistance, immune escape, etc. However, challenges remain in capturing complex mutation patterns and supporting diverse viral families with varying genome architectures. To address these needs, we present \href{https://variant.up.railway.app/}{VARIANT}, an openly accessible web server for mutational analysis of RNA viral genomes and associated viral products across both single- and multi-segment virus genomes. The server takes as input a viral reference genome, a reference protein sequence, and/or multiple sequence alignment, and automatically provides full annotation of mutation types, including standard categories such as point mutations (missense, silent, and nonsense), insertions, deletions, or frameshift events in both coding and non-coding regions. In addition, VARIANT detects three biologically significant mutation patterns that often overlooked by conventional software/packages: ``\textit{row mutations}'' (consecutive substitutions within a window of 3 nucleotides), ``\textit{hot mutations}'' (two non-consecutive substitutions within a window of 3 nucleotides), and potential \textit{programmed ribosomal frameshifting} (PRF) regions. The server currently contains automatic analysis of major viral pathogens, including SARS-CoV-2, HIV-1, Influenza H3N2, Ebola virus, and Chikungunya virus. It also allows users to analyze additional viruses by uploading custom genome information. Users can track VARIANT analysis progress in real time, visualize mutation distributions, and download structured results in ZIP format. VARIANT also incorporates dual graph topology analysis to classify frameshifting element structures from dot-bracket notation input. This feature enables systematic comparison of RNA secondary structure motifs across viral families by mapping structures to a comprehensive library of dual graph topologies. The web server is freely available at \href{https://variant.up.railway.app/}{https://variant.up.railway.app/}.
\end{abstract}
\blfootnote{$\dagger$: Corresponding Author}

% {\bf Keywords:} Viral Mutation, Frameshifting, Webserver
\newpage
%\pagenumbering{roman}
%\begin{verbatim}
%\end{verbatim}
%
% {\setcounter{tocdepth}{4} \tableofcontents}
%%
 \newpage
 %\clearpage
 %\pagebreak

\setcounter{page}{1}
\renewcommand{\thepage}{{\arabic{page}}}
% \begin{multicols}{2}
% \multicollinenumbers
% \linenumbers

% \begin{multicols}{2}
% \multicollinenumbers
% \linenumbers

\section{Statement of Significance}
    Understanding how viruses mutate is crucial for controlling pandemics, developing effective treatments, and designing vaccines. However, existing computational tools often cannot identify important mutation patterns or analyze the structural features that regulate viral protein production, such as frameshifting elements (FSEs), through programmed ribosomal frameshifting. VARIANT fills these gaps by providing the first comprehensive web platform that simultaneously detects overlooked mutation patterns, annotates mutations across diverse viral genome architectures, and classifies the RNA structural topologies of frameshifting elements using dual graph approaches. By packing these functions into a web interface, VARIANT empowers researchers to perform sophisticated viral genomic analyses in an intuitive way. This web server can accelerate research in viral evolution, drug resistance mechanisms, and antiviral therapeutic development across multiple viral families.

\section{Introduction}
Rapid and accurate characterization of viral mutations is essential for understanding viral evolution, transmission dynamics, drug resistance, and immune escape mechanisms. For example, delayed identification of emerging mutations hindered rapid public health responses during the COVID-19 pandemic, resulting in wide spread of variants with increased transmissibility or immune evasion.
Similar challenges arise in chronic viral infections such as HIV-1, where mutation-driven drug resistance continually impacts treatment effectiveness, and in rapidly evolving RNA viruses such as influenza, Ebola, and Chikungunya. Therefore, easy-to-access state-of-the-art computational tools to decode mutational landscapes in viral genomes and associated protein products are needed. 

Several mutation analysis tools/web-servers have been developed, each addressing a portion of this need. \href{https://gisaid.org/database-features/covsurver-mutations-app/}{CoVsurver}, integrated within the \href{https://gisaid.org/}{GISAID} platform, annotates mutations in SARS-CoV-2 genomes and marks notable variants. However, its functionality is restricted to the submitted sequences in GISAID and does not apply to other viral species. The Nextclade, part of the Nextstrain system, aligns viral genomes to curated references and identifies nucleotide substitutions and small indels \cite{aksamentov2021nextclade}. While widely used for SARS-CoV-2, influenza, and several other pathogens, Nextclade primarily focuses on per-site mutations and quality control (QC) metrics. For instance, clusters of adjacent substitutions are typically flagged as potential sequencing errors (SNP-cluster QC) rather than interpreted as biologically meaningful mutation patterns. Besides web-based resources, variant callers such as V-Phaser 2 \cite{yang2013vphaser} and LoFreq \cite{wilm2012lofreq} excel at identifying intrahost low-frequency variants from deep sequencing reads, yet they do not offer genome- or protein-level functional annotation.

Despite their utility, these tools do not address several problems. First, most platforms prioritize single-nucleotide variants and small indels, leaving limited support for systematic annotation of coding consequences (missense, silent, and nonsense mutations), non-coding mutations, or frameshift-generating indels across genomic sequences. Second, biologically meaningful combinatorial mutation patterns, such as non-adjacent substitutions or consecutive substitutions that preserve reading frames, are generally ignored or mislabeled as artifacts. These patterns could indicate localized editing or compensatory evolutionary events. Third, many tools are specialized for a single virus (e.g., SARS-CoV-2) or require manual configuration to support new pathogens, limiting their utility for comparative or cross-virus analyses. Critically, existing tools lack systematic approaches to characterize RNA secondary-structure topology in functionally critical regions, such as programmed ribosomal frameshifting elements, which are essential regulatory features in many RNA viruses.

To address these limitations, we present VARIANT, a publicly accessible web server for comprehensive analysis of viral mutations at both the genome and protein levels. VARIANT supports both single-segment and multi-segment RNA viruses and provides automated annotation of all major mutation categories, including point mutations (missense, silent, nonsense), insertions, deletions, frameshift events, and mutations in non-coding regions. A unique feature of VARIANT is its ability to detect two biologically significant but frequently overlooked mutation patterns: row mutations (consecutive substitutions within a three-nucleotide window) and hot mutations (non-consecutive clustered substitutions within short genomic regions). VARIANT further identifies genomic regions with signatures of programmed ribosomal frameshifting (PRF), which play essential roles in several RNA viruses such as HIV-1 and coronaviruses. Additionally, VARIANT integrates dual graph topology analysis to classify RNA secondary structures of frameshifting elements, enabling systematic comparison of structural motifs across viral families.

The VARIANT server currently includes built-in support for major viral pathogens, including SARS-CoV-2, HIV-1, Influenza H3N2, Zaire ebolavirus, and Chikungunya virus. In addition, VARIANT allows users to analyze other species by uploading custom reference genomes and pre-aligned multiple sequence alignments. Users can track processing jobs in real time, visualize mutation distributions, and download results in structured CSV or ZIP formats. Together, these capabilities make VARIANT a unified and extensible platform for the viral genomics community, enabling mutation interpretation, evolutionary analysis, and downstream studies in vaccine design, antibody design, antigen design, antiviral drug development, antigenic analysis,  and viral surveillance.

\section{Materials and Methods}
\subsection{The Overview of the VARIANT Workflow}
VARIANT is a publicly accessible web server designed to provide comprehensive mutation analysis for viruses, integrating genome- and protein-level annotations, specialized pattern detection, and support for both single- and multi-segment viruses. The platform is developed based on previously published techniques  \cite{yin2020genotyping,wang2020mutations} and is implemented in Python with a modular architecture that separates user interaction, mutation processing, and the result output into distinct components. It is implemented via Railway and uses the FastAPI framework to support asynchronous task execution, real-time progress tracking, and structured file management.

\begin{figure}[ht!]
    \includegraphics[width=1.0\textwidth]{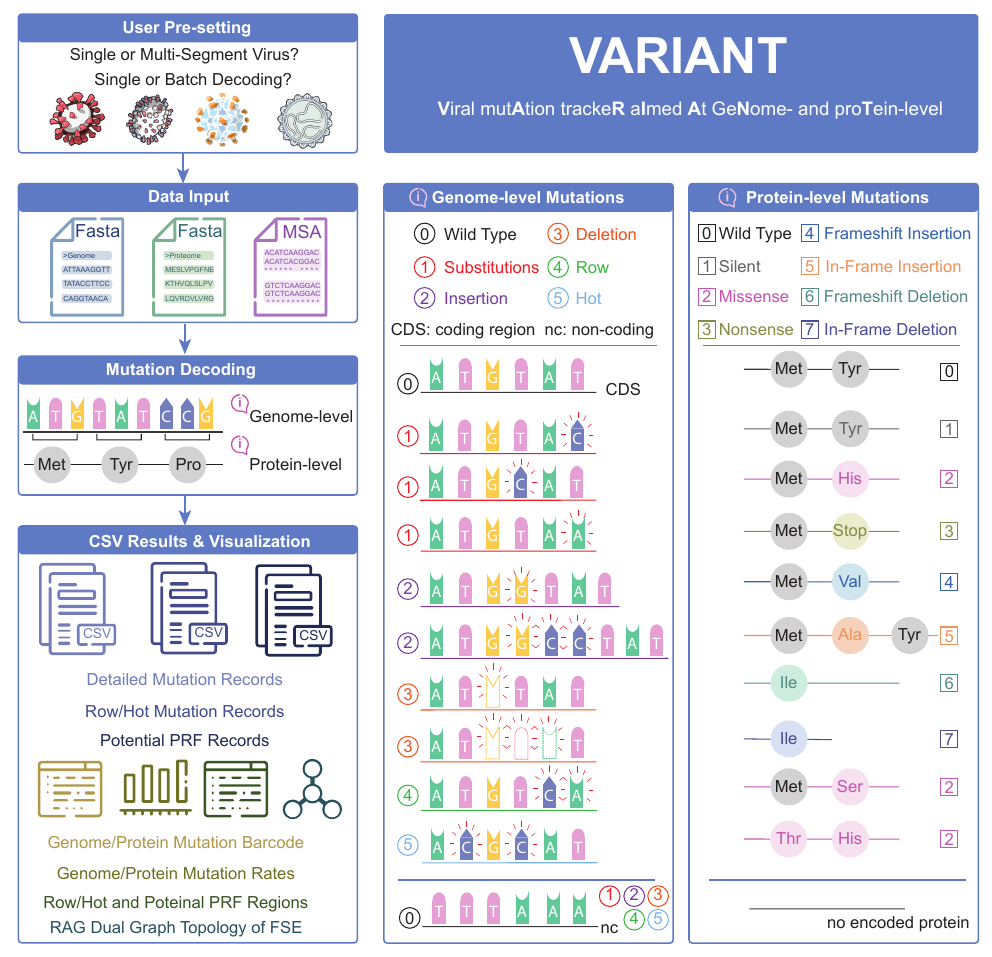}
    \centering
    \caption{Overview of the VARIANT web server for genome- and protein-level viral mutation analysis. The VARIANT web server supports both single- and multi-segment viral genomes and offers single or batch decoding modes. The workflow begins with user pre-configuration and input of reference genome and proteome sequences (FASTA Format) along with aligned viral genomes (MSA Format). The server performs mutation decoding at both the nucleotide (genome-level) and amino acid (protein-level) resolution, supporting standard mutation types such as substitutions, insertions, and deletions, as well as complex mutation patterns including row mutations (consecutive substitutions within a window of 3 nucleotides), hot mutations (two non-consecutive substitutions within a window of 3 necleotides), and potential programmed ribosomal frameshift (PRF) signals. Additionally, VARIANT incorporates dual graph topology analysis to classify RNA secondary structures of frameshifting elements using the "RNA-As-Graphs" framework. Output includes structured CSV files of detailed mutation records, row/hot mutation sites, and potential PRF regions. Visualizations of mutation distributions and classification summaries are also provided. Mutation classification is shown on the right, including genome-level categories and their corresponding protein-level consequences.}
    \label{fig:workflow}
\end{figure}

The typical VARIANT workflow (see \autoref{fig:workflow}) begins with three required input files from users: 1) a reference genome in FASTA format, 2) a corresponding reference proteome (amino acid sequences), and 3) a multiple sequence alignment (MSA) file of viral genomes aligned to the reference. Users can either select a pre-configured virus (e.g., SARS-CoV-2, HIV-1, Influenza H3N2, Chikungunya, Zaire Ebola) or upload custom virus reference files (including reference genome, reference protein sequence, and MSA) to analyze other viral species. Once inputs are provided, the server automatically performs nucleotide-level variant calling\footnote{Here, `variant calling' means the identification of all sequence differences at the nucleotide level relative to the reference genome.} across all aligned genomes in the MSA, maps mutations to their corresponding codons and protein positions, and classifies mutation types, including silent, missense, nonsense, insertions, deletions, frameshift, and non-coding region mutations.

In addition to standard mutation annotations, the server performs automated detection of three complex and biologically relevant mutation patterns: 1) row mutations, which are defined as two or three consecutive substitutions within a 3-nucleotide window; 2) hot mutations, which are defined as two non-consecutive substitutions within a 3-nucleotide window; and 3) potential programmed ribosomal frameshifting (PRF) sites, which are identified by the presence of known slippery sequences and downstream RNA structure motifs. 

Upon completion, VARIANT provides users with structured outputs, including genome mutation annotation files, aggregated protein-level summaries, row/hot mutation records, and bundled ZIP archives for streamlined data management. The web interface is fully automated and allows users to monitor the analysis pipeline in real time. Once users click the "Generate All Visualizations", the interactive visualizations are provided to further interpret the results, such as genome-wide mutation barcode maps, mutation rate statistics on each protein, row/hot mutation spots, and potential PRF sites.

\subsection{Mutation Detection and Classification}
Following multiple sequence alignment (MSA) input, VARIANT performs comprehensive genome-wide mutation detection by comparing each aligned genome sequence to the specified viral reference genome. This comparison identifies all nucleotide differences, which are categorized into standard (substitution, insertion, deletion) and non-standard mutation (row and hot mutation) types. The mutation detection pipeline is applied to each genome in the alignment (excluding the reference) and supports both single-segment and multi-segment viral genomes.

\subsubsection{Standard Mutation Types on the Coding Region}
\autoref{tab:standard_mutation_cds} summarizes three standard mutation types observed in the coding regions of RNA viral genomes, along with their corresponding protein-level mutations. A {\it substitution} mutation, also known as point mutation, occurs when a single nucleotide in a codon of the reference genome is replaced by a different nucleotide. If this altered codon encodes the same amino acid, the mutation is {\it silent} (or {\it synonymous}). If it encodes a different amino acid, the mutation is {\it missense}, while the mutation is {\it nonsense} if it creates a premature stop codon. 

{\it Insertion} mutations involve the addition of one or more nucleotides into the genome. When the number of inserted nucleotides is a multiple of three, the reading frame remains intact and the protein gains one or more amino acids (an in-frame insertion). However, if the insertion is not divisible by three, the reading frame shifts result in a {\it frameshift} mutation that typically alters all downstream codons and often produces a nonfunctional or truncated protein.

{\it Deletion} mutations represent the removal of one or more nucleotides from the genomic sequence. When the number of deleted nucleotides is a multiple of three, the reading frame remains intact, resulting in an in-frame deletion that removes one or more amino acids from the protein product. If the deletion is not divisible by three, it disrupts the reading frame and causes a {\it frameshift deletion}.

\begin{table}[htbp]
\centering
\caption{Mapping of standard genome-level mutations (on the coding region) to corresponding protein-level effects.}
\label{tab:standard_mutation_cds}
\begin{tabular}{>{\centering\arraybackslash}m{4.0cm}>{\centering\arraybackslash}m{4.5cm}m{6.0cm}}
\toprule
\textbf{\makecell{Genome-Level Mutation}} & \textbf{\makecell{Definition}} & \textbf{Possible Protein-Level Consequences} \\
\midrule
\textbf{\makecell[cc]{Substitution}} & \makecell{One nucleotide is replaced} & 
1) Silent (synonymous) \newline 
2) Missense (amino acid (AA) change) \newline 
3) Nonsense (introduces stop codon) \\
\midrule
\textbf{\makecell{Insertion}} & \makecell{Insertion nucleotide(s)} & 
1) In-frame insertion (adds AAs) \newline
2) Frameshift insertion (disrupts the reading frame, truncated protein) \\
\midrule
\textbf{\makecell{Deletion}} & \makecell{Delete nucleotide(s)} & 
1) In-frame deletion (removes AAs) \newline
2) Frameshift deletion (alters AA sequence and likely truncation) \\
\bottomrule
\end{tabular}
\end{table}

\subsubsection{Non-standard Mutation Types on the Coding Region}
A core innovation of VARIANT is its ability to move beyond standard mutation cataloging to detect and characterize complex and biologically relevant mutation patterns that are often overlooked by existing mutation parser software. VARIANT includes specialized approaches for detecting two mutation patterns: {row mutations} and {hot mutations}. These patterns are of particular interest in viral evolution and adaptation studies, where clustered or coordinated nucleotide substitutions may signal mutational hotspots, (AID/APOBEC)-mediated deamination \cite{li2024exploring}, or co-evolving residues under selective pressure.

A {\it row mutation} is defined as a set of two or three consecutive nucleotide substitutions within a 3-nucleotide window. This pattern captures short runs of adjacent mutations that may arise from replication slippage, APOBEC/ADAR-mediated editing, or compensatory substitutions that preserve structural or functional properties of the encoded protein. Unlike standard point mutation reporting, which treats such changes as independent events, VARIANT identifies and records row mutations as unified mutational blocks.

VARIANT defines {\it hot mutations} as two non-consecutive substitutions within a sliding 3-nucleotide window, where the middle nucleotide remains unchanged. This pattern is indicative of clustered but independent mutation events localized within a short region of the genome. Hot mutations may reflect immune pressure, epitope targeting, or error-prone replication hotspots. VARIANT scans each genome sequence using an overlapping window approach and flags all qualifying hot mutation sites.

Both row and hot mutations are reported in dedicated CSV files with detailed annotations including the affected genomic window, substituted bases, and flanking context. These records enable downstream analysis of mutational clustering patterns and their association with functional or structural genome elements. By systematically detecting and annotating these patterns, VARIANT provides researchers with a higher-resolution view of mutational dynamics beyond isolated single-nucleotide variants.

\begin{table}[htbp]
\centering
\caption{Mapping of non-standard genome-level mutations (on the coding region) to corresponding protein-level effects.}
\label{tab:non-standard_mutation_cds}
\begin{tabular}{>{\centering\arraybackslash}m{4cm} >{\centering\arraybackslash}m{4.5cm}m{6cm}}
\toprule
\textbf{\makecell{Genome-Level Mutation}} & \textbf{\makecell{Definition}} & \textbf{Possible Protein-Level Consequences} \\
\midrule
% \textbf{Multi-nucleotide variant (MNV)} & Two or more consecutive nucleotide substitutions. & 
% \textbullet\ Missense (typically) \newline 
% \textbullet\ Nonsense (if stop codon is introduced) \\
\textbf{Row mutation} & Two or three consecutive substitutions within a 3-nt window & 
1) Silent (synonymous) \newline 
2) Missense (amino acid change) \newline 
3) Nonsense (introduces stop codon)  \\
\midrule
\textbf{Hot mutation} & Two non-consecutive substitutions within a 3-nt window & 
1) Silent (synonymous) \newline 
2) Missense (amino acid change) \newline 
3) Nonsense (introduces stop codon) \\
\bottomrule
\end{tabular}
\end{table}

\subsubsection{Mutations on the non-coding region}
VARIANT tracks all types of mutations mentioned in \autoref{tab:standard_mutation_cds} and \autoref{tab:non-standard_mutation_cds} within non‑coding regions of viral genomes. %Although there is no direct amino acid change on the non-coding region, they may affect regulation, splicing, transcription, and translation. 
Although such mutations do not directly alter amino acid sequences, they may have significant functional consequences, such as affecting RNA splicing or gene regulation. For example, substitutions or indels in untranslated regions (UTRs) can disrupt regulatory elements, influence transcript stability, or alter mRNA secondary structures, thereby modulating translation efficiency or RNA degradation pathways.

\subsection{Detection of Potential $+$1/$-$1 PRF Regions} 
Programmed ribosomal frameshifting (PRF) is a translational recoding mechanism that allows ribosomes to pause and shift reading frames during protein synthesis, resulting in the production of alternative protein products from a single mRNA transcript. PRF mechanisms include: $-$1 PRF and $+$1 PRF, depending on the direction of the frameshift. In $-$1 PRF, the ribosome shifts one nucleotide backward (in the 5' direction), while in $+$1 PRF, it shifts one nucleotide forward (in the 3' direction). 

\subsubsection{$-$1 Programmed Ribosomal Frameshifting ($-$1 PRF)}
Many viruses, including retroviruses, alphaviruses, and coronaviruses, exploit $-$1 PRF to synthesize essential fusion proteins and maintain proper gene expression ratios. The canonical $-$1 PRF signal comprises three elements: 1) a slippery heptanucleotide sequence of the form X\_XXY\_YYZ (where the triplet XXX includes identical nucleotides, YYY is usually AAA or UUU, and Z is generally any base except G); a well-known example is UUUAAAC found in HIV and coronaviruses \cite{newton2025conformational,yan2025heterogeneous}; 2) a short spacer ($\thicksim$ 5-9 nucleotides) separating the slippery site from the downstream structure, which ensures proper geometric positioning of the ribosome during the pause; and 3) a stable downstream RNA secondary structure, often a stem-loop or pseudoknot, which stalls the ribosome and increases the probability of simultaneous slippage of the P-site and A-site tRNAs into the $-$1 frame. The $-$1 PRF is used to regulate the synthesis of essential fusion proteins such as Gag–Pol in HIV and the ORF1a/ORF1b polyproteins in coronaviruses.

\subsubsection{$+$1 Programmed Ribosomal Frameshifting ($+$1 PRF)}
The $+$1 PRF mechanism is less common than $-$1 PRF, but functionally important in specific cellular and viral contexts. It is typically stimulated when the ribosome encounters an inefficiently decoded codon in the A-site (due to low tRNA abundance or reliance on wobble base pairing), which promotes realignment of tRNAs in the $+$1 frame. In many cases, $+$1 PRF signals are coupled to the presence of an in-frame stop codon in the 0 frame; the frameshift allows the ribosome to bypass termination and continue elongation. Unlike $-$1 PRF, $+$1 PRF does not always require a strong downstream pseudoknot, although nearby secondary structures or context-specific sequences can contribute to ribosomal pausing. 
% A well-characterized example is the mammalian OAZ1 (ornithine decarboxylase antizyme) mRNA, where a +1 frameshift ensures synthesis of the full-length regulatory protein. 
Overall, $+$1 PRF highlights the role of codon context, tRNA availability, and ribosomal pausing as critical determinants of frameshift efficiency.

\subsubsection{Workflow for the Detection of $-$1 and $+$1 PRFs}
The detection of $-$1 and $+$1 programmed ribosomal frameshifting (PRF) follows a structured computational workflow that integrates sequence, structural, and translational features. The process begins by scanning genomic sequences for canonical slippery heptamers (e.g., X\_XXY\_YYZ) that promote ribosomal slippage either one nucleotide backward ($-$1 PRF) or forward ($+$1 PRF). For each identified site, the tool evaluates spacer lengths between 5 and 9 nucleotides, as these distances critically affect frameshifting efficiency.

Next, the program predicts potential RNA secondary structures downstream of slippery sites using three complementary tools: 
\begin{enumerate}
\item \textbf{RNAfold}, an algorithm in the ViennaRNA package \cite{lorenz2011viennarna} which identifies stem-loops and calculates minimum free energy (MFE);
\item \textbf{ProbKnot} \cite{bellaousov2010probknot}, a method which detects pseudoknots using heuristic methods;
\item {\textbf{PKNOTS} \cite{rivas1999dynamic}, an algorithm which applies a dynamic programming approach to resolve pseudoknot motifs;}
\item \textbf{NUPACK} \cite{dirks2004algorithm}, a 2D folding program that can predict simple pseudoknots.
\end{enumerate}

For RNA secondary-structure prediction with pseudoknot detection, we provided results from three packages: ProbKnot\cite{bellaousov2010probknot}, PKNOTS\cite{rivas1999dynamic}, and NUPACK\cite{dirks2004algorithm} so that users can choose the prediction most suitable for their analysis. These tools rely on distinct algorithmic frameworks and offer complementary strengths in identifying the RNA structures essential for PRF induction. Within each package, structures with lower minimum free energy (MFE) values are interpreted to be more stable. However, we caution that the fold space of FSEs is complex and involves multiple sequence-length dependent folds. See references \cite{yan2025heterogeneous,yan2022lengthdependent,lee2026kinetic,portillo2026open} for example. Thus, our tools are meant as starting research points for various FSE investigations. 

An optional tRNA-based module assesses the translational context. For each candidate site, it computes tRNA availability at the A- and P-site codons, evaluates ribosomal pausing potential, and identifies wobble base pairs (e.g., G--U) that impair decoding. This module generates an overall tRNA score: high tRNA abundance lowers frameshifting likelihood, whereas wobble interactions enhance it.

The final output integrates genomic coordinates, RNA structural features, spacer lengths, frame context (if GFF/GTF annotations are provided), and tRNA metrics. By combining sequence motifs with RNA structure and codon context, VARIANT offers a comprehensive and biologically informed strategy to identify PRF candidates in functional genomics and translational control studies.

It is important to note that our pipeline reports all candidate PRF-like signals, so identification of a site does not guarantee that frameshifting occurs in vivo. Experimental validation remains essential for confirming the functional relevance of predicted sites.

\subsection{RNA Secondary Structure Topology Classification via RNA-As-Graphs}
Programmed ribosomal frameshifting (PRF) efficiency is believed to depend on the secondary structure formed downstream of the slippery sequence, commonly referred to as the frameshifting element (FSE). The FSE acts as a structural barrier that impedes ribosomal translocation, thereby increasing the probability of frameshifting. Distinct FSE topologies, ranging from simple stem-loops to more complex pseudoknots, are associated with different frameshifting efficiencies and often display lineage-specific distributions across viral families \cite{yan2022lengthdependent}.

To enable systematic comparison of FSE topologies across viral species and putative PRF candidates, VARIANT incorporates a dual-graph classification pipeline based on the RNA-As-Graphs (RAG) framework developed by the Schlick laboratory (See \href{https://github.com/Schlicklab/Existing-Dual-Search}{Existing-Dual-Search} at GitHub) \cite{gan2003exploring}. In this framework, RNA secondary structures with pseudoknots are represented as two-dimensional dual graphs as follows: each double-stranded helical stem with at least 2 base pairs is encoded as a vertex. Single-stranded regions connecting secondary elements such as bulges, loops, and junctions are encoded as edges. The 5' and 3' termini are omitted from explicit representation. Because this graph-based abstraction captures the global fold architecture independently of sequence identity and helix length, it enables identification of topologically equivalent RNA structures despite substantial sequence divergence. The coarse graining also reduces conformational space size and makes many analysis, classification \cite{wang2025how}, and design problems \cite{jain2020inverse} more approachable.

The classification in VARIANT accepts RNA secondary structures in dot-bracket notation as input. For users starting with experimentally determined 3D structures, secondary structures can be extracted from PDB or mmCIF files using the \href{https://rnapdbee.cs.put.poznan.pl/}{RNAPdbee 3.0} web server \cite{pielesiak2026rnapdbee}, which annotates secondary structures of knotted and unknotted RNAs and outputs dot-bracket notation. Alternatively, users can directly input pre-computed dot-bracket notation from structure prediction tools (e.g., NUPACK, IPknot, PKnots) or from the RNAPDBee 3.0 database.

The RAG dual graph classification progresses in three stages. First, RNA secondary structures in dot-bracket notation (with filename extension as .dbn) are normalized to Dissecting the Spatial Structure of RNA (DSSR) records that preserve multi-strand information and pseudoknot annotations using \href{https://rnapdbee.cs.put.poznan.pl/}{RNAPdbee 3.0}. Second, these normalized records are converted into connectivity table (CT) format using the \texttt{dot2ct} utility from the RNAstructure package \cite{mathews2006rna}. Third, CT files are processed using \texttt{dualGraphs.py} from the \href{https://github.com/Schlicklab/Existing-Dual-Search}{Existing-Dual-Search} repository to extract dual graph representations and perform topology matching against the dual graph library (\texttt{Dual\_Library.txt} in \href{https://github.com/Schlicklab/Existing-Dual-Search}{Existing-Dual-Search}). Here, \href{https://github.com/Schlicklab/Existing-Dual-Search}{Existing-Dual-Search} repository was originally developed by the Schlick laboratory and integrated here into VARIANT for topology classification \cite{zhu2022rnaasgraphs}. In this step, dual-graph topologies are characterized using graph invariants, including eigenvalue spectra and degree-based connectivity features, and are then matched against a curated reference library via a two-step procedure consisting of spectral filtering followed by exact adjacency-matrix isomorphism testing. Structures that satisfy both criteria are assigned canonical dual-graph identifiers (graph IDs) and visualized using the matplotlib/plotly library. 

On the VARIANT web server, users can submit RNA secondary structures in dot-bracket notation as input, and the platform returns the corresponding dual graph identifiers together with graphical representations of the assigned RNA topologies.

\section{Results and Discussion}
\subsection{Case Studies for Genome-Level and Protein-level Mutation Analysis}
\subsubsection{Standard Mutation Visualization and Analysis}
\begin{figure}[ht!]
    \includegraphics[width=1.0\textwidth]{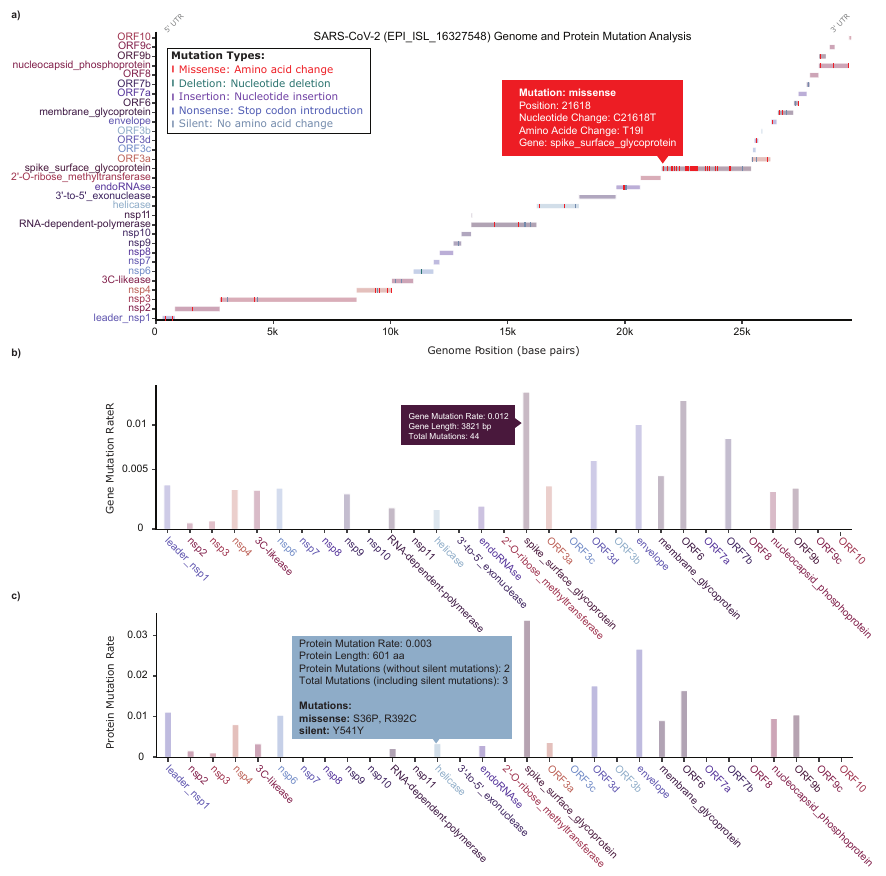}
    \centering
    \caption{Genome organization and distribution of mutations across SARS-CoV-2 viral genes. {\bf a)} The top panel shows the viral genome as horizontal bars representing individual genes/proteins. Mutations are displayed as vertical lines at their respective genomic positions. Here, different colors represent different mutation types (red: missense, green: deletion, purple: insertion, blue: nonsense, grey: frameshift). The $x$-axis indicates genomic position in base pairs, and the $y$-axis lists gene names. Hover interactions reveal detailed mutation information, including nucleotide changes, protein changes, and genomic coordinates. This web visualization enables rapid identification of mutation hotspots and gene-specific mutation patterns across the viral genome. {\bf b)} Gene-level mutation rate, calculated as the total number of nucleotide changes normalized by gene length. Each bar corresponds to a gene and is colored consistently with the top panel. The $y$-axis shows mutation rate (mutations per base pair), and the $x$-axis lists gene names. Hover interactions display additional details, including mutation counts and gene lengths. {\bf c)} Protein-level mutation rate (silent mutation is not counted), calculated as the number of total amino acid changes normalized by protein length. Each bar represents a protein, with hover interactions revealing protein length and detailed mutation records.}
    \label{fig:sars-cov-2_summary}
\end{figure}

VARIANT employs an interactive visualization framework powered by \href{https://plotly.com/python/}{Plotly}. As shown in \autoref{fig:sars-cov-2_summary} {\bf a)}, the VARAINT visualization tool presents a mutation barcode aligned with viral gene regions. Each mutation is represented as a vertical line positioned at its exact genomic coordinate on a horizontal gene schematic. Colors distinguish 5 mutation types: red for missense, green for deletions, purple for insertions, blue for nonsense, and grey for silent mutations. The $x$‑axis indicates genomic position in base pairs, while gene identifiers are displayed along the $y$‑axis. Interactive hover functionality enriches the interface. Users can access detailed metadata, including nucleotide changes, corresponding amino acid alterations, mutation type, etc. This panel enables intuitive identification of mutational hotspots and gene‑specific mutation patterns, assisting researchers in pinpointing areas of potential functional importance at a glance. \autoref{fig:sars-cov-2_summary} {\bf b)} and {\bf c)} are barplots for gene-level and protein-level mutation rates. A longer bar indicates a higher mutation rate. Here, gene-level mutation rate is defined as the total number of nucleotide changes normalized by gene length, and the protein-level mutation rate is calculated as the number of total amino acid changes normalized by protein length (silent mutation is not counted). This analysis reveals which genomic regions/proteins accumulate mutations at the highest rate, which provides insights into evolutionary pressure and functional consequences. 

\subsubsection{Non-standard Mutation Visualization and Analysis}
\begin{figure}[hbt!]
    \includegraphics[width=1.0\textwidth]{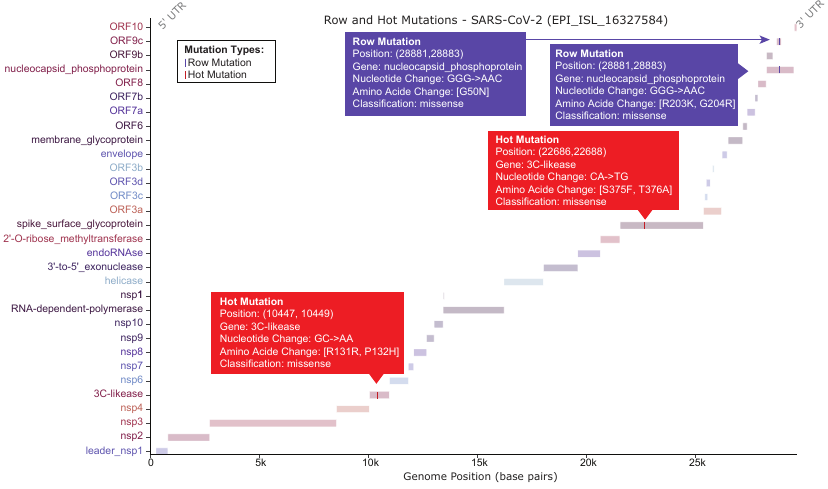}
    \centering
    \caption{SARS-CoV-2 Genome organization with specialized row and hot mutation pattern detection. The viral genome is displayed as horizontal bars, each representing a distinct gene (for a single-segment virus) or segment (for a multi-segment virus). Row mutation (purple color) and hot mutation (red color) patterns are overlaid as vertical lines at their corresponding genomic positions. The $x$-axis indicates genomic position in base pairs, while the $y$-axis lists gene names. Interactive hover features provide detailed information on each mutation, including mutation type, genomic position, nucleotide and amino acid changes, and biological classification. }
    \label{fig:sars-cov-2_rowhot}
\end{figure}

\autoref{fig:sars-cov-2_rowhot} illustrates the row and hot mutations on a specific genome region. Row mutation (purple color) and hot mutation (red color) patterns are overlaid as vertical lines at their corresponding genomic positions. This analysis enables the detection of complex mutational signatures that may have distinct biological implications beyond those of single-point mutations. For example, there are three consecutive mutations G28881A, G28882A, G28883C, which result in two amino acid mutations on nucleocapid (R203K and G204R). This triplet of mutations has been found in over 60\% of SARS-CoV-2 sequences and is now nearly fixed in the global population \cite{mears2025emergence}, which creates a new transcription regulatory sequence site that leads to the creation of a new open reading frame called N.iORF3. N.iORF3 has been shown to be transcribed during infection, indicating a new functional protein being produced by the virus.

\subsubsection{Programmed Ribosomal Frameshifting (PRF) Detection Analysis}
To evaluate the performance of $-$1 PRF detection pipeline, we apply it to four well-characterized viral genomes that present $-$1 PRF: SARS-CoV-2, HIV-1, Ebola virus, and Chikungunya virus. Each of these viruses utilizes $-$1 PRF to regulate essential protein expression during translation.

For SARS-CoV-2, VARIANT identifies the $-$1 PRF site at the ORF1a/1b junction, centered on the slippery sequence UUUAAAC (site starts at 13462) and a downstream RNA pseudoknot. This frameshift enables synthesis of the ORF1ab polyprotein \cite{yan2022lengthdependent}.

For HIV-1, the pipeline detects the $-$1 PRF site at the Gag-Pol junction, marked by the heptameric slippery sequence UUUUAAG and a predicted downstream stem-loop. This event regulates production of the Gag-Pol fusion protein, which is critical for reverse transcription and viral assembly.

For Ebola virus, which employs a less canonical PRF mechanism, the pipeline identified candidate signals within the VP40 matrix protein region. These align with evidence that VP40 isoforms arise via frameshifting, although regulatory features remain less defined than in retroviruses or coronaviruses.

For Chikungunya virus, a predicted $-$1 PRF event was found within the structural polyprotein region, consistent with the proposed elements modulating capsid-to-envelope protein stoichiometry. The identified site and downstream structure support a functional PRF mechanism in alphavirus translation.

\subsection{Dual Graph Topology Analysis of Viral Frameshifting Elements}
\begin{figure}[hbt!]
    \includegraphics[width=1.0\textwidth]{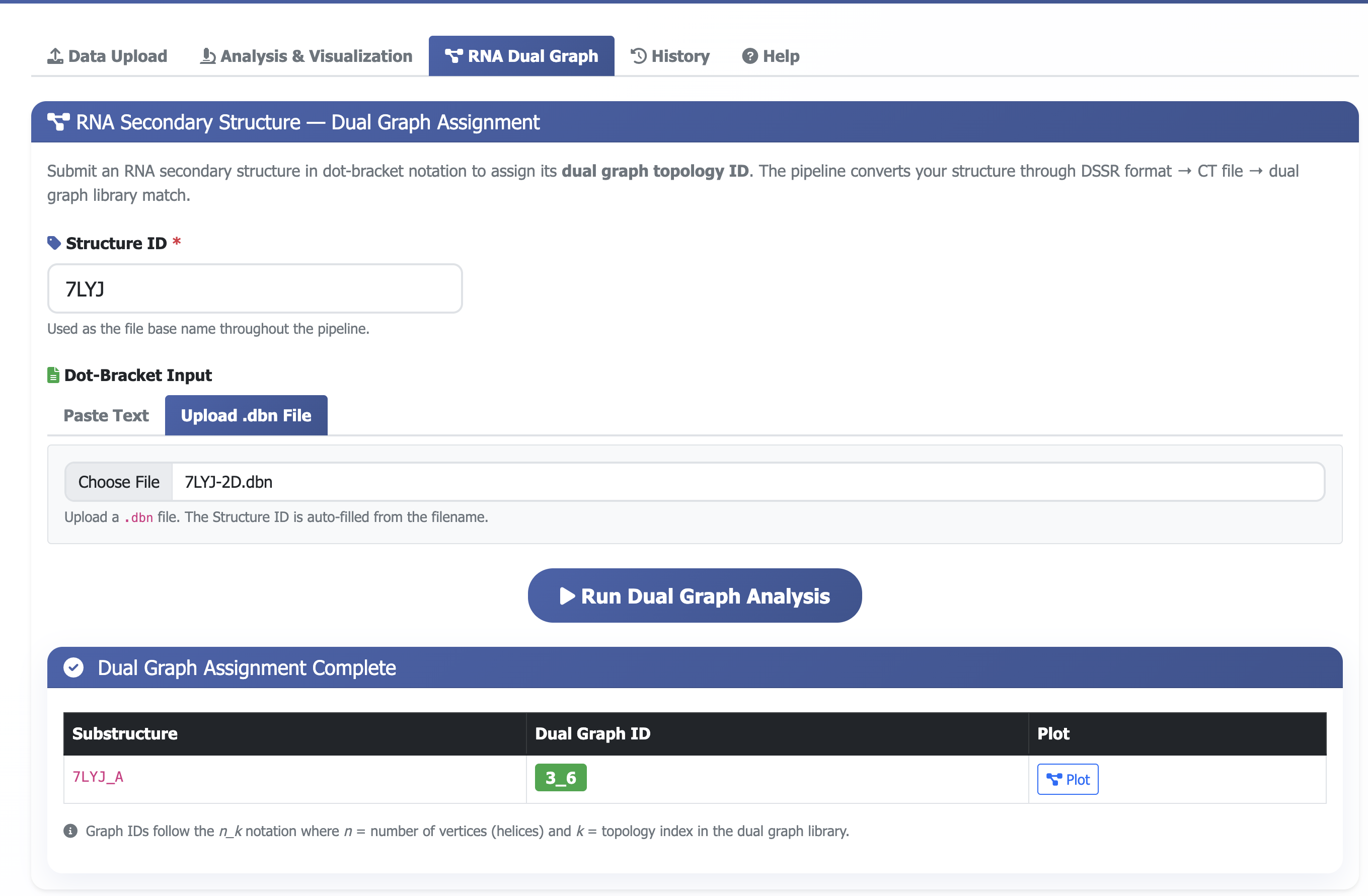}
    \centering
    \caption{VARIANT web interface for dual graph analysis. The input panel accepts dot-bracket notation from structure prediction tools, and the output display presents the dual graph ID and its corresponding visual representation. The example shows the SARS-CoV-2 frameshifting element (PDB ID: 7LYJ), which exhibits a pseudoknot motif dual graph 3\_6.}
    \label{fig:web dual graphs}
\end{figure}

As described above, VARIANT provides a platform to convert RNA secondary structures into their dual graph representations for viral FSEs. Users can input dot-bracket notation obtained from secondary structure prediction tools such as NUPACK, IPknot, PKnots, or RNAPDBee, and VARIANT outputs the corresponding dual graph identifier along with representative graph motifs from our updated dual graph library, as shown in the \autoref{fig:web dual graphs}.

The web interface streamlines the analysis workflow by automatically matching input structures against our comprehensive dual graph library. For example, when analyzing the SARS-CoV-2 frameshifting element (PDB ID: 7LYJ), VARIANT identifies its dual graph motif and provides the structural classification, enabling rapid comparison with other frameshifting elements in our database. This topological representation reveals conserved architectural features that may be critical for frameshifting function.

\begin{figure}[ht!]
    \includegraphics[width=1.0\textwidth]{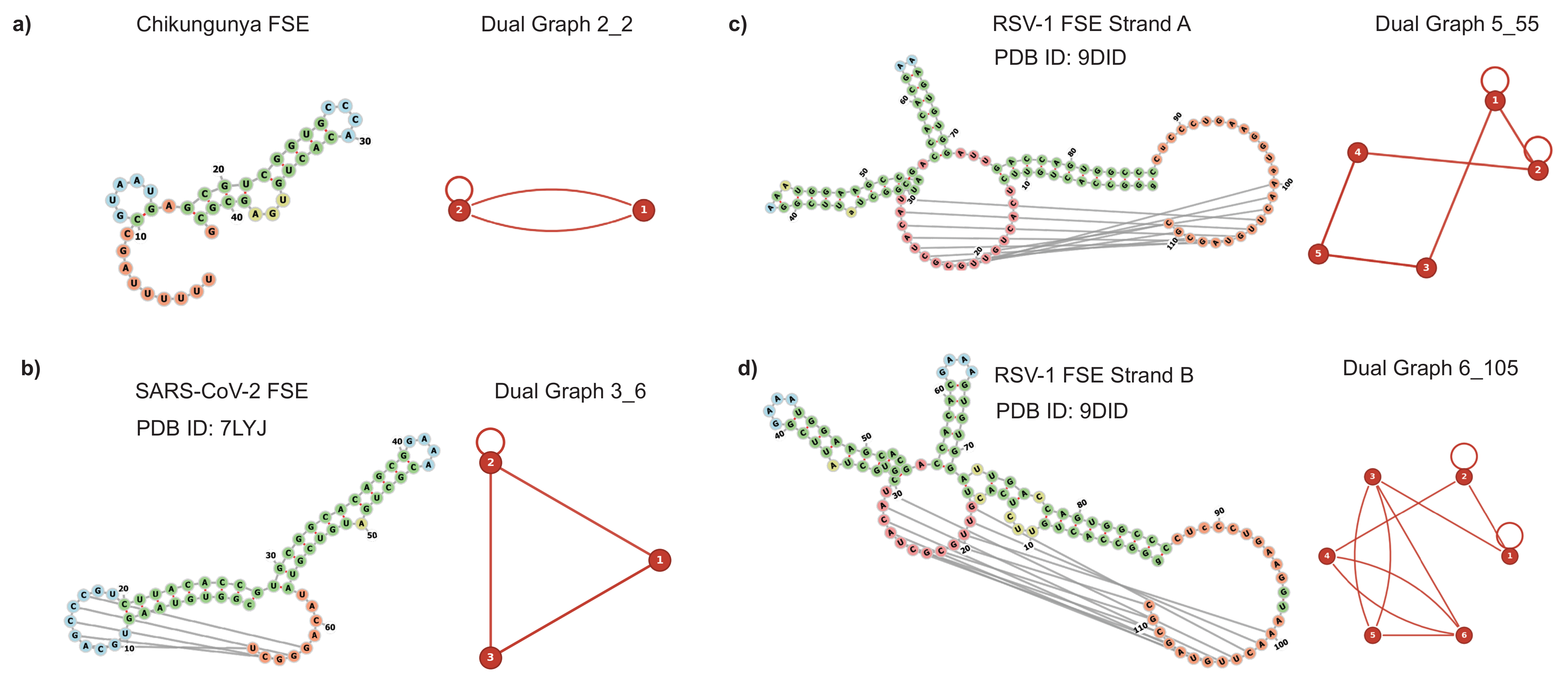}
    \centering
    \caption{Dual graph representations of viral frameshifting elements. Each row represents a different frameshifting element: {\bf a)} Chikungunya virus frameshifting element \cite{lee2026kinetic}, {\bf b)} SARS-CoV-2 frameshifting pseudoknot RNA \cite{yan2022lengthdependent}, {\bf c)} Rous sarcoma virus frameshifting pseudoknot RNA strand A \cite{jones2025structural}, and {\bf d)} Rous sarcoma virus frameshifting pseudoknot RNA strand B \cite{jones2025structural}. For each structure, the secondary structure diagram (left) produced from forna \cite{kerpedjiev2015forna} and dual graph ID (right) are shown. }
    \label{fig:fse dual graphs}
\end{figure}

We applied this analysis to a curated set of viral frameshifting elements from multiple virus families, including the Chikungunya virus frameshifting element \cite{lee2026kinetic}, SARS-CoV-2 frameshifting pseudoknot RNA \cite{yan2022lengthdependent}, and Rous sarcoma virus frameshifting pseudoknot RNA \cite{jones2025structural}.  \autoref{fig:fse dual graphs} presents the dual graph representations of representative frameshifting elements. Each panel shows the secondary structure (left) and its corresponding dual graph ID (right). All dot-bracket notations are obtained via the RNAPDBee 3.0 website \cite{pielesiak2026rnapdbee}. Notably, structures containing pseudoknots exhibited distinct dual graph patterns characterized by multiple edges connecting the same pair of vertices, representing the cross-linked loops that form the pseudoknot structure. In dual graph representation, pseudoknots create cyclic paths where vertices (helical stems) are connected by edges (loops) in topologies that cannot be represented by tree graphs \cite{yan2022lengthdependent}. This topological classification provides a quantitative framework for comparing frameshifting elements across different viral species and identifying structural determinants of frameshifting efficiency.

\section{Conclusion}
We developed VARIANT as a comprehensive web platform for analyzing mutations in RNA viral genomes. The server addresses key limitations in existing tools by providing unified support for diverse viral species, detecting biologically meaningful mutation patterns that are typically overlooked, and integrating structural analysis of programmed ribosomal frameshifting elements through "RNA-As-Graph" dual graph identification.

VARIANT automates the complete annotation pipeline from nucleotide changes to protein-level consequences, handling both standard mutations (substitutions, insertions, deletions) and complex patterns such as row mutations, hot mutations, and frameshift events.  In addition, the integration of dual-graph topology analysis represents a novel feature in viral mutation analysis platforms. By classifying frameshifting element structures using the RNA-As-Graphs framework, VARIANT enables direct comparison of RNA secondary structure motifs across different viral lineages. This capability bridges sequence-level mutation analysis with structural characterization.

VARIANT currently supports five major viral pathogens with pre-configured analysis pipelines, while remaining extensible to additional viruses through custom genome uploads. The platform's modular architecture, real-time progress tracking, and interactive visualizations make complex viral genomics analysis accessible to researchers.

\section{Data and Software Availablity}
The VARIANT web server is freely available at \href{https://variant.up.railway.app/}{https://variant.up.railway.app/} and requires no user registration. The source code of \href{https://variant.up.railway.app/}{VARIANT}  is available at \href{https://github.com/wangru25/VARIANT}{https://github.com/wangru25/VARIANT}. The original code of dual graph detection and visualization is available at \href{https://github.com/Schlicklab/Existing-Dual-Search}{https://github.com/Schlicklab/Existing-Dual-Search}.

\section{Author Contributions}
Rui Wang (Conceptualization, Methodology, Software, Writing - original draft, Writing - review \& editing), Xuhang Dai (Software, Validation, Writing - review \& editing), Xin Cao (Software, Validation, Writing - review \& editing), Changchuan Yin (Conceptualization, Methodology, Software, Validation, Supervision, Writing - review \& editing), Tamar Schlick (Conceptualization, Methodology, Supervision, Writing - review \& editing), Guo-Wei Wei (Conceptualization, Methodology, Supervision, Writing - review \& editing)

\section{Acknowledgments}
This work was supported in part by NIH Grants R01AI164266 and R35GM148196, NSF Grants DMS-2052983 and IIS-1900473, MSU Research Foundation, Bristol-Myers Squibb 65109, and Georgia Research Alliance to G-W. W. Support to T.S. from the National Institutes of Health, National Institute of General Medical Sciences Award R35-GM122562, National Science Foundation Awards (DMS-215177 and DMS-2330628) from the Division of Mathematical Sciences, and Philip-Morris USA Inc. is gratefully acknowledged. This work was also supported by a grant from the Simons Foundation through the NYU Simons Center for Computational Physical Chemistry Award MPS-T-MPS-00839534 to MET. X.D. acknowledges partial support from a graduate fellowship from the Simons Center for Computational Physical Chemistry (SCCPC) at NYU.

\newpage
% % \bibliographystyle{abbrv}
\bibliographystyle{unsrt}
% %% \bibliographystyle{custom}
% \bibliography{VARIANT}
\bibliography{refs}
% % \end{multicols}

\end{document}